\documentclass[12pt]{iopart}
\usepackage{bm}
\usepackage{epsfig}
\usepackage{iopams}
\usepackage{amssymb}
\usepackage{color}

\newcommand{\ee}{\mathrm{e}}

\begin{document}

\title[Emergence of particle clusters]
      {Emergence of particle clusters  in a one-dimensional model: connection to condensation processes}
\author{Matthew Burman}
\author{Daniel Carpenter}
\author{Robert L. Jack}
\address{Department of Physics, University of Bath, Bath BA2 7AY, United Kingdom}

\begin{abstract}
We discuss a simple model of particles hopping in one dimension with attractive interactions.  Taking a hydrodynamic limit in which the interaction strength increases with the system size, we observe the formation of multiple clusters of particles, with large gaps between them.  These clusters are correlated in space, and the system has a self-similar (fractal) structure.  These results are related to condensation phenomena in mass transport models and to a recent mathematical analysis of the hydrodynamic limit in a related model.
\end{abstract}

%\maketitle

\section{Introduction}

A familiar theme in statistical mechanics is that particles interacting by simple dynamical rules can lead to complex emergent behaviour -- familiar examples include the rich phenomenology of fluid dynamics and turbulence which appear generically when atoms or molecules interact by momentum-conserving collisions, or the wide range of thermodynamic phases that are available for spherical (isotropically-interacting)  particles.  Even in much simpler model systems such as exclusion processes or zero-range processes in one dimension, unexpected phenomena continue to surprise physicists and mathematicians, including condensation~\cite{evans-review,majumdar05,evans-multi,chleboun14,cates98,gross11,waclaw12} and unusual fluctuation phenomena~\cite{bod04,bertini,harris06,hurtado14,jack-HU}.

Here, we investigate a very simple model with unusual condensation behavior.  We consider $N$ particles that diffuse in a one-dimensional periodic system of size $L$.  The particles are coupled to a heat bath at temperature $T$ and interact via attractive forces from their nearest neighbours, which leads to the formation of clusters of particles.  We focus on a \emph{hydrodynamic limit} of large system size $L$, in which the particle density $\rho=N/L$ is fixed, but the interaction strength increases in the limit.   We find that the particles self-organise into a large number of clusters. The number of particles in each cluster diverges in the limit; at the same time, each cluster becomes concentrated on a single point.  The relation to condensation is that the large gaps between clusters correspond to the kinds of condensate that appear in mass transport models~\cite{majumdar05}, following a mapping described in~\cite{cates98,evans-review}.  The unusual feature of the model considered here is that the system forms many large clusters or, equivalently, many condensates.   Systems with multiple condensates have been investigated before~\cite{evans-multi}, but this effect is much less studied than systems with a single condensate, and the mechanism of condensation in our case differs from~\cite{evans-multi}.  There are also similarities between this work and the traffic flow model of~\cite{cates98}, where each cluster considered here would correspond to a traffic jam.  {However, the system considered here has an equilibrium steady state (which is symmetric under time reversal), so the clusters necessarily move diffusively (without any preferred direction).}

As well as offering a new twist on condensation, this work is also motivated by connections between this model system and a recent mathematical study~\cite{max} where it was found that the dynamics of the particle density in a similar model should be described by a stochastic partial differential equation (PDE) with a stochastic term that does not vanish in the hydrodynamic limit.  Usually, one expects to recover (almost surely) deterministic behaviour in the hydrodynamic limit: for example, the deterministic diffusion equation describes the spreading of a large number of random walkers.  
 Moreoever, the stochastic PDE found in~\cite{max} is closely related to the Dean equation~\cite{dean}, which describes (in this case) the diffusive motion of a finite number of non-interacting particles.  This result offers the possibility that the clusters that form in our model might themselves act as free particles that diffuse through the system.  However, the arguments of~\cite{max} do not provide a simple physical picture of the behaviour of the underlying particle model.  By exploring its behaviour in more detail, we find that the emergence of clusters is consistent with the existence of a finite stochastic element to the dynamics even in the hydrodynamic limit.   However, these clusters do not diffuse as free particles, but are instead rather strongly interacting, leading to a scale-invariant distribution of clusters within the system.  We argue that an understanding of the hydrodynamic limit of this model requires an understanding of the dynamics of the clusters that form in the system -- this work establishes a foundation for future work in that area.

%Second, the model is connected (by a standard mapping [?cite]) to a model of mass transport, in which a given mass of material is distributed across $N$ discrete sites, and moves between them according to a random \emph{chipping process}~\cite{majumdar05}.  In such models, there are a range of well-known condensation phenomena, in which a finite fraction of the total mass ends up localised either on a single site, or (in some cases) on several sites.  Our model results in a similar phenomenon -- the large gaps between the clusters in the particle model map to macroscopically-occupied sites in the transport model.  However, the condensation that we observe is of an unusual type: in the limit that we consider, there are many condensates.  A similar situation arises in~\cite{evans-multi}, but the correlations among the condensates are quite different in this case.

The structure of the paper is as follows.  In Sec.~\ref{sec:model} we introduce the model.  In Sec.~\ref{sec:static} we derive some basic results for its static (equilibrium) properties, including the existence of an instability towards cluster formation at a finite temperature $T^*$, and its behaviour in the thermodynamic limit.  
In Sec.~\ref{sec:multi}, we consider the limit in which multiple macroscopic clusters appear and we analyse the (non-trvial) structure of this state.  Finally in Sec.~\ref{sec:conc} we discuss the implications of these results and their connection to previous work.

\section{Model}
\label{sec:model}

\newcommand{\EE}{\mathcal{E}}

The model consists of $N$ particles that move in a one dimensional system of size $L$, with periodic boundaries.  The position of particle $i$ is $x_i \in [0,L)$.  Let the distance between particle $i$ and the nearest particle to its right be $y_i$.  Each particle interacts only with its nearest left and right neighbours so the energy of the system can be written in the form
\begin{equation}
E(\bm{x}) = \sum_i \EE(y_i) ,
\label{equ:Ey}
\end{equation}
where we introduced the vector $\bm{x}=(x_1,x_2,\dots,x_N)$, from which the gap sizes $\{y_i\}$ can be calculated.

We focus on the specific case $\EE(y_i)=J\log y_i$ with $J>0$, so that particles feel attractive forces from their neighbours.  In this case, the energy can take arbitrarily large negative values when one or more gaps are very small.  To  avoid theoretical difficulties associated with this effect, it is sometimes convenient to regularise the energy, for which we consider two possibilities: we can either take $\EE(y_i)=J \log \max(y_i,\epsilon)$ for some small constant $\epsilon$, or we  give each particle a hard core of size $\epsilon$, so that $\EE(y_i)=+\infty$ if $y_i<\epsilon$.  We are primarily interested in the behaviour as $\epsilon\to0$: we believe that all the results that we present here are valid in that limit {(independent of the choice of regularisation scheme)}.

We consider this system to be coupled to a heat bath at temperature $T$ so that, given sufficiently long time, we expect the system to equilibrate.  In that case, the probability (or probability density) of finding the system in configuration $\bm{x}$ is given by
a Boltzmann distribution, 
\begin{equation}
P(\bm{x}) = \frac{1}{Z_x(T)} {\rm e}^{-E(\bm{x})/T} ,
\label{equ:boltz}
\end{equation}
 where $Z_x(T)$ is a normalisation constant (partition function); {we work throughout in units where Boltzmann's constant $k_{\rm B}=1$.}  The formula (\ref{equ:boltz}) assumes that this distribution is normalisable, which is certainly true for any $\epsilon>0$ but may fail for $\epsilon=0$: we return to this question below.

{Within this system, the density $\rho=N/L$ sets the only natural length scale.  For a given number of particles $N$, the behaviour of the model depends on two dimensionless parameters, which are the (dimensionless) inverse temperature $\beta=J/T$ and regularisation parameter $\epsilon_0 = \epsilon\rho$.}

\subsection{Dynamical evolution}

We consider two dynamical rules for the evolution of the model in time.  In the first, the particles evolve according to a Langevin equation as
\begin{equation}
\gamma\partial_t x_i =  -\frac{\partial E}{\partial x_i} + \sqrt{2\gamma T} \eta_i ,
\label{equ:lang}
\end{equation}
where $\gamma$ is a friction constant (which acts only to set the units of time), $\eta_i$ is a Gaussian distributed white noise with zero mean and $\langle \eta_i(t) \eta_i(t')\rangle = \delta(t-t')$.  This choice is simple from a theoretical point of view and is consistent with~\cite{max}, but the model is difficult from a numerical perspective because of the large values of ${\partial E}/{\partial x_i}$ that appear when particles approach one another.  In the absence of any interparticle forces ($E=0$), the single particle diffusion constant is $D_0=T/\gamma$.
{Since the energy depends logarithmically on the particle separations, such processes are related to the motion of particles in logarithmic potentials, which appears in several contexts in physics, as discussed in~\cite{hirschberg}.}

The second choice, which is more convenient numerically, uses a Monte Carlo (MC) dynamics to evolve the system according to a Markov chain.  The method depends on a parameter $a_{\rm max}$, which is the maximum displacement of a single particle in a single MC move.  In each MC move, a particle $i$ is chosen at random, and a displacement $\Delta x$ is chosen uniformly from $[-a_{\rm max},a_{\rm max}]$. The particle is moved from $x_i$ to $x_i+\Delta x$ and the change in energy associated with this move is calculated.  This move is accepted with a probability given by the Metropolis formula $p_{\rm acc}=\min(1,\ee^{-\Delta E/T})$ where $\Delta E$ is the change in energy associated with the move.  If the move is not accepted then the particle is returned to its original position $x_i$.  After each attempted move, the time is incremented by  $a_{\rm max}^2/(6D_0N)$ so that in the absence of interparticle forces, the diffusion constant for the MC dynamics matches that of the Langevin equation (\ref{equ:lang}).

\begin{figure}
\hspace{2cm}\includegraphics[width=8cm]{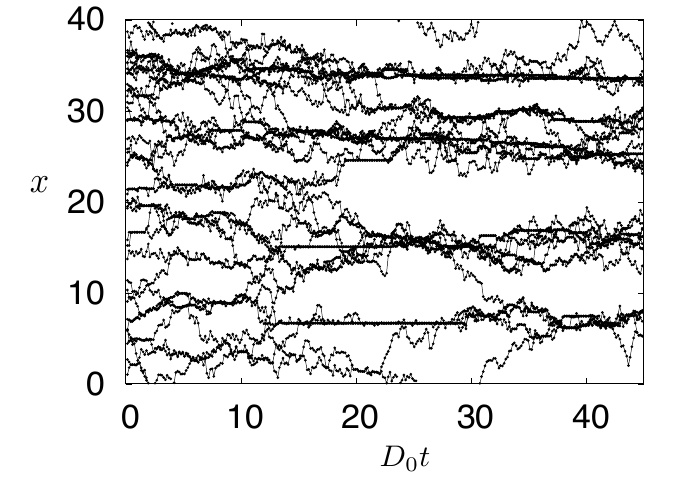}
\caption{A trajectory of the system with $N=40$ particles at density $\rho=1$, with $\beta=0.75$ and $a_{\rm max}=0.3$.  Particles are initially distributed at random in the system but as time progresses, clusters of particles are observed to form.}\label{fig:traj}
\end{figure}

Note that within the MC method, the ordering of the particles within the system may change, since particles are free to ``overtake'' each other.  Also, the system respects detailed balance with respect to the equilibrium distribution {(\ref{equ:boltz})} so, for large times, the system should converge to that distribution.  Moreover, in the limit $a_{\rm max}\to0$ (assuming now that $\epsilon>0$), this dynamical MC method converges to the solution of the Langevin equation (\ref{equ:lang}).  However, we note that the results presented here are far from the limit $a_{\rm max}\to0$, in particular, this limit may require $a_{\rm max}\ll \epsilon$ while our numerical results have $a_{\rm max}\gg \epsilon$.  Nevertheless, we emphasise that the steady state distribution of the system is given by (\ref{equ:boltz}), independent of $a_{\rm max}$ (as long as the distribution (\ref{equ:boltz}) is normalisable).

All numerical results in this work are obtained with the MC dynamics.  Fig.~\ref{fig:traj} shows a trajectory of the system, at inverse temperature $\beta=0.75$.  At $t=0$ the particles are distributed at random, but one clearly sees that they self-organise into clusters.  Of course this result is expected since particles can reduce their energy by approaching each other.  The questions that we address in the following relate to this cluster formation. 

\section{Static (equilibrium) properties}
\label{sec:static}

In order to investigate cluster formation, it is useful to consider the distribution (probability density) for the gap sizes $y_i$.  We consider the model with $\epsilon=0$, in which case (\ref{equ:boltz}) leads to
\begin{equation}
P(\bm{y}) = \frac{1}{Z_y(\beta)} \delta\left(L-\sum_{i=1}^N y_i\right) \prod_{i=1}^N y_i^{-\beta} ,
\label{equ:y2}
\end{equation}
where $\bm{y}=(y_1,y_2,\dots,y_N)$; also $\beta=J/T$ is the (dimensionless) inverse temperature, the function $\delta(x)$ is a Dirac delta, and $Z_y(\beta)$ is the partition function for this representation of the system.  Distributions of this form are familiar from zero-range processes and from mass transport models~\cite{evans-review,majumdar05,cates98,chleboun14}.

One sees from (\ref{equ:y2}) that this probability density diverges as $y_i\to0$, and that the distribution will not be normalisable for $\beta\geq 1$.  In fact, $\beta=1$ (or $T=J$) is a special temperature for this model, as we now discuss.
%
%We will see later that as the system is cooled ($\beta\to 1^-$ or $T\to J^+$), the mean energy approaches $-\infty$, signalling that the regularisation parameter $\epsilon$ can no longer be neglected, and that this parameter controls the behaviour of the system.  For that reason, the analysis in this article is restricted to $\beta<1$ (that is, high temperatures, $T>J$), in which case the limit $\epsilon\to0$ is smooth and (\ref{equ:y2}) is valid.  We briefly discuss the behaviour for $\beta>1$ in Sec.~\ref{subsec:chip}, below.
%
%\emph{ put this later --}
%Note that the models considered in~\cite{majumdar05} lead to probability distributions similar to (\ref{equ:y2}), but in those cases condensation occurs only for $\beta>2$ (i.e., at a significantly lower temperature than we consider here).  The difference is that the energy function $\cal E$ associated with those models is bounded below by a constant of order unity and they do not take any limit of small $\epsilon$.  In that case, condensation occurs due to singular behaviour coming from large values of $y$: in the case considered here, this singularity comes from \emph{small} $y$.
%
It is useful to calculate the marginal distribution of a single gap within this system, which is
\begin{equation}
P_1(y_1) = \int_{[0,L]^{N-1}}\! \mathrm{d}y_2\dots\mathrm{d}y_N\, P(\bm{y}) .
\end{equation}
Making the change of variables $\tilde{y}_i = y_i/(L-y_1)$ for $i=2\dots N$ yields
\begin{eqnarray}
\fl
P_1(y_1) & = & \frac{(L-y_1)^{(N-1)(1-\beta)-1}}{Z_y(\beta)} y_1^{-\beta}  \int_{[0,a]^{N-1}}\! \mathrm{d}\tilde y_2\dots\mathrm{d}\tilde y_N\, 
\delta\left(1-\sum_{i=2}^N \tilde{y}_i\right) \prod_{i=2}^N \tilde{y_i}^{-\beta} ,
\label{equ:p1y1-int}
%\nonumber \\
%& = & \frac{1}{Z_y(\beta)} (\dots) \int_{[0,1]^{N-1}}\! \mathrm{d}\tilde y_2\dots\mathrm{d}\tilde y_N\, 
%\delta(1-\sum_i \tilde{y}_i) \prod_i \tilde{y_i}^{-\beta}
%\nonumber \\
%& = & C^{-1} y_1^\beta (L-y_1)^{N-2}
\end{eqnarray}
where $a=L/(L-y_1)$.  The Dirac delta constrains all integration variables to be less than or equal to unity and we have $a>1$, so the integration domain $[0,a]^{N-1}$ can be replaced by $[0,1]^{N-1}$.  The resulting integral is independent of $y_1$ so (at least for $\beta<1$) it can be absorbed into the normalisation constant, yielding
\begin{equation}
P_1(y) = \frac{1}{C_1} y^{-\beta} (L-y)^{(N-1)(1-\beta)-1},
\label{equ:p1y-finite}
\end{equation}
where $C_1$ is a normalisation constant.
Hence the rescaled gap length $y/L$ follows a Beta distribution.  
It will be useful in the following to recall the definitions of three special functions
\begin{eqnarray}
\Gamma(u) & := & \int_0^\infty t^{u-1} {\rm e}^{-t} \mathrm{d}t ,
\nonumber \\
\psi(u) & := & \frac{1}{\Gamma(u)} \frac{\mathrm{d}}{\mathrm{d}u} \Gamma(u) ,
\nonumber \\
B(u,v) & := & \int_0^1 t^{u-1} (1-t)^{v-1} \mathrm{d}t = \frac{\Gamma(u)\Gamma(v)}{\Gamma(u+v)} ,
%\nonumber \\
%\frac{\partial}{\partial u}B(u,v) & = &  B(u,v) [ \psi(u) - \psi(u+v) ]
\end{eqnarray}
which are the Gamma function ($\Gamma$), the digamma function ($\psi$) and the Beta function ($B$).
Using these results, 
normalisation of $P_1(y)$ means that for $\beta<1$ (high temperature) we have
$ C_1=L^{-\beta+(N-1)(1-\beta)} B(1-\beta,(N-1)(1-\beta))$. 

\subsection{Low temperature behaviour and effects of regularisation}

{For $\beta\geq1$, the distribution $P_1(y)$ in (\ref{equ:p1y-finite}) is not normalisable: one sees that $P_1$ diverges at $y=0$ in such a way that its integral does not exist (there is also another non-integrable divergence at $y=L$).  Physically, the source of this problem is that small gaps $y_i$ lead to unbounded negative contributions to the energy of the system, and corresponding divergences in the distribution $P(\bm{y})$ given in (\ref{equ:y2}).  For $\beta>1$, these divergences are strong enough that $P(\bm{y})$ cannot be normalised, and so cannot be interpreted as a probability density any more.  

To understand this effect, we regularise the energy $\cal E$ as described in Sec.~\ref{sec:model} so that particles may not approach each other more closely than a distance $\epsilon$. 
In this case we define a regularised distribution $P^\epsilon(\bm{y})$ by replacing $\prod_i y_i^{-\beta}$ in (\ref{equ:y2}) with $\prod_i y_i^{-\beta}\Theta(y_i-\epsilon)$ where $\Theta$ is a Heaviside (step) function.  We also replace the partition function $Z_y(\beta)$ by $Z_y^\epsilon(\beta)$.
(The following arguments are easily generalised to the alternative regularisation in which particles may approach each other arbitrarily closely, but with the energy for small gaps being bounded below.)

The partition function for the regularised model is
\begin{eqnarray}
\fl
Z_y^\epsilon(\beta) = \int_{[\epsilon,L]^N} \mathrm{d} y_1\dots\mathrm{d} y_N\, 
\delta\left(L-\sum_{i=1}^N {y}_i\right) \prod_{i=1}^N {y_i}^{-\beta} 
\nonumber \\
= \int_{[\epsilon,L]^{N-1}} \mathrm{d} y_2\dots\mathrm{d} y_N\, 
\left(L-\sum_{i=2}^N {y}_i\right)^{-\beta} \Theta\left(L-\epsilon-\sum_{i=2}^N {y}_i\right) \prod_{i=2}^N {y_i}^{-\beta} .
\label{equ:zeps}
\end{eqnarray}
The final integrand is bounded above by $\epsilon^{-N\beta}$ and the integration range is finite, so the integral always exists.  Hence the distribution $P^\epsilon(\bm{y})$ for this regularised model is normalisable, and it can be interpreted as a probability density.  This indicates that the regularisation does indeed make the model well-defined.  The remaining question is whether (or under what circumstances) the regularisation parameter $\epsilon$ can be chosen small enough that the relevant physical observables in the model do not depend on $\epsilon$.  

We defer a rigorous analysis of the small-$\epsilon$ limit to a later work.  For our purposes, observe that if  $\lim_{\epsilon\to0} Z_y^\epsilon(\beta)=Z_y(\beta)$ then for any $\bm{y}$ we have $P^\epsilon(\bm{y})\to P(\bm{y})$ in (\ref{equ:y2}).  Physical observables in the system are calculated as averages with respect to $P^\epsilon(\bm{y})$: if the observable of interest is bounded in magnitude then this analysis is sufficient to ensure that it converges to a finite limit as $\epsilon\to0$.  In that case one can always choose $\epsilon$ small enough that the regularisation has no significant effect.  We will show that this is the case whenever $\beta<1$.  On the other hand, if $Z_y^\epsilon$ diverges as $\epsilon\to0$ then clearly $P^\epsilon$ does not converge to the function $P$ in  (\ref{equ:y2}).  This is the case for $\beta\geq 1$.

We first consider $\beta>1$. {The integrand in (\ref{equ:zeps}) is non-negative so one may obtain a lower bound on the integral by replacing the range $[\epsilon,L]^{N-1}$ by $[\epsilon,A]^{N-1}$ for any $A<L$.  Restrict  $\epsilon$ to be smaller than some $\epsilon^*$ and fix some constant $A$ in the interval $(\epsilon^*,\frac{L-\epsilon^*}{N-1})$. In this case the step function in the integrand of (\ref{equ:zeps}) is equal to unity throughout the integration domain. Finally, note that $(L-\sum_{i=2}^N {y}_i)^{-\beta} \geq L^{-\beta}$. Combining all the ingredients yields $Z_y^\epsilon(\beta) \geq \int_{[\epsilon,A]^{N-1}} L^{-\beta} \prod_{i=2}^{N} y_i^{-\beta} {\rm d}y_2\dots {\rm d}y_N= L^{-\beta}\left(\frac{\epsilon^{1-\beta}-A^{1-\beta}}{\beta-1}\right)^{N-1}$.  For $\beta>1$, this bound diverges as $\epsilon\to0$ so $Z^\epsilon_y(\beta)$ diverges in this limit, and the behaviour of the model depends strongly on the regularisation parameter $\epsilon$ even when this parameter is small.  A similar effect is observed for $\beta=1$: in that case the divergence is logarithmic in $\epsilon$.  The origin of this divergence is again the diverging probability for small gaps that renders the distribution $P_1$ in (\ref{equ:p1y-finite}) non-normalisable.}

%The integrand in (\ref{equ:zeps}) is always greater than $\prod_{i=2}^N {y_i}^{-\beta}$ from which we clearly see that if $\beta>1$ then $Z^\epsilon_y(\beta) \geq (\frac{\epsilon^{-(\beta-1)}-L^{-(\beta-1)}}{\beta-1})^{N-1}$.  Hence $Z^\epsilon(\beta)$ diverges as $\epsilon\to0$, so the behaviour of the model depends strongly on the regularisation parameter even when $\epsilon$ is small.  A similar effect is observed for $\beta=1$: in that case the divergence is logarithmic in $\epsilon$.  The origin of this divergence is again the diverging probability for small gaps that renders the distribution $P_1$ in (\ref{equ:p1y-finite}) non-normalisable. 

For $\beta<1$, the integral in (\ref{equ:zeps}) can be evaluated directly for $\epsilon=0$: one substitutes $y_N=y'_N(L-\sum_{i=2}^{N-1} y_i)$ which allows the $y_N$ integral to be performed (yielding a Beta function); one then repeats the same procedure to perform the integration over $y_{N-1}$, and so on.  No divergences appear so we conclude that $Z^\epsilon_y$ does indeed have a finite limit as $\epsilon\to0$, and the probability distribution $P^\epsilon(\bm{y})\to P(\bm{y})$ for $\epsilon\to0$, as asserted above. Note that in the following we sometimes consider limits where $\beta\to1^-$: in such cases one should always take $\epsilon\to0$ before any limit of $\beta\to 1$.}

\subsection{Mean energy and mean gap size}

We also calculate
the mean energy per particle which (in units of $J$) is $\frac{1}{NJ} \langle E\rangle = \langle \log y_i \rangle = \int_0^L\mathrm{d}y P_1(y) \log y $, where the angle brackets denote averages in the equilibrium state of the system.  Writing $\log y = \lim_{\delta\to0} \frac{1}{\delta} (y^\delta-1)$ and using properties of the Beta function yields the energy per particle
\begin{equation}
\frac{1}{NJ} \langle E \rangle = \psi(1-\beta) - \psi(N(1-\beta)) + \log L .
\label{equ:enN}
\end{equation}
%Note that for $\beta<1$, one may take the thermodynamic limit, $N,L\to\infty$ at fixed $\rho=N/L$, and the energy per particle is finite (** give the value) because $\psi(x)\sim \log x$ at large $x$.  
{The digamma function diverges for $x\to0$ as $\psi(x)\simeq -1/x$, so taking $\beta\to1^-$, we have $\frac{1}{NJ} \langle E \rangle \simeq (-1+N^{-1})/(1-\beta)$.} That is, the energy becomes large and negative in this limit, again signalling that the system is unstable and small gaps are predominating.

Finally, it is useful to consider the average fraction of the system that is taken up by gaps with sizes between $y$ and $y+\mathrm{d}y$, which is $P_{\rm g}(y) \mathrm{d}y$, with 
\begin{equation}
P_{\rm g}(y) = \rho y P_1(y) .
\label{equ:pg}
\end{equation}
Compared with $P_1(y)$, the main feature of this distribution is that while there may be very many gaps with small $y$, these take up only a small fraction of the system.  For $0<\beta<1$, this means that $P_{\rm g}(y)$ tends to zero as $y\to0$, in contrast to $P_1(y)$ which diverges.  If one picks a random point in the system then the size of the gap containing this point is distributed as $P_{\rm g}$, and the mean of this distribution is easily verified to be
\begin{equation}
\overline{Y}_{\rm g} = \frac{\langle y_i^2 \rangle}{\langle y_i\rangle} = \frac{L(2-\beta)}{1+N(1-\beta)} .
\label{equ:Yg}
\end{equation}
(Note that $\sum_i y_i=L$ independent of the arrangement of the particles, so one always has $\langle y_i\rangle=L/N=1/\rho$, but the values of $\langle y_i^2\rangle$ and $\overline{Y}_{\rm g}$ are sensitive to the structure of the system.  In the following we sometimes refer to $\overline{Y}_{\rm g}$ as the ``mean gap size'': we note that this is the mean associated with $P_{\rm g}$, which is different from $\langle y_i \rangle$ because each gap is weighted by its size within the distribution $P_{\rm g}$.)

\begin{figure}
\includegraphics[width=15cm]{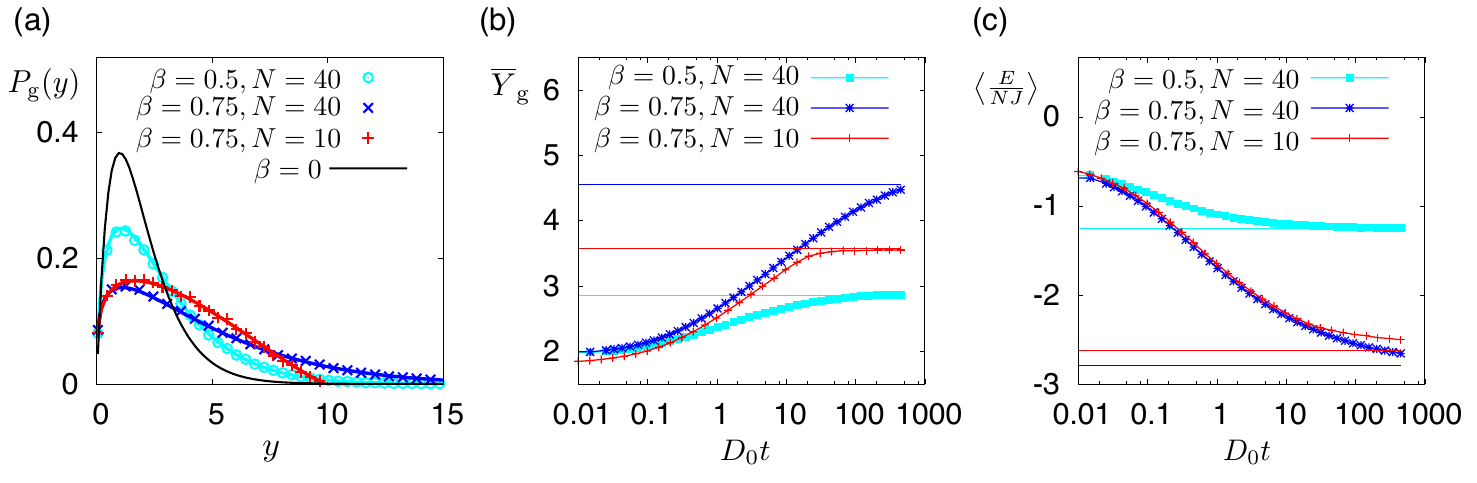}
\caption{Numerical results for finite systems, for $\rho=N/L=1$ and $a_{\rm max}=0.3$.  We show two temperatures $\beta=0.5,0.75$ and for the lower temperature we show results for two system sizes $N=10,40$.  (a) Equilibrium gap size distributions $P_{\rm g}(y)$ obtained numerically (points) and compared with theoretical predictions (solid lines).  The numerical results were obtained at time $D_0 t=450$.  (b)~Time evolution of the mean gap size $\overline{Y}_{\rm g}$, with equilbrium values shown as solid horizontal lines. (c)~Time evolution of the average energy per gap $\langle E \rangle/(NJ)$.  This quantity is sensitive to the gap size distribution at small $y$ -- for the lower temperature, this quantity has not fully converged even for the largest times considered, since the MC dynamics used are not efficient for sampling very small gaps.}
\label{fig:finiteN}
\end{figure}

Figure~\ref{fig:finiteN} shows results obtained with the MC dynamics, and compared with the theoretical predictions of this section.  For these calculations we work at unit density $\rho=N/L=1$ with $a_{\rm max}=0.3$ (fixing the density simply fixes the unit of length since the interaction potential $\cal E$ has no characteristic length scale).  Within the numerical calculations, we regularise using a very small value of $\epsilon$, of the order of the machine precision: the results do not depend on the precise value of $\epsilon$ and they agree with the predictions for the limit $\epsilon\to0$.  We interpret this as evidence that the limit $\epsilon\to0$ is regular for the dynamics as well as for the equilibrium properties, as long as $\beta<1$ (that is, $T>J$). The gap size distribution $P_{\rm g}(y)$ agrees well with the theoretical predictions.  For this distribution, the most apparent effect of the attractive forces between particles is to enhance the probability of \emph{large gaps} -- this is due to the formation of clusters of particles, with large gaps between them.
 
Starting from a random initial condition, Fig.~\ref{fig:finiteN} also shows the convergence to equilibrium of the mean gap size $\overline{Y}_{\rm g}$ and the mean energy per particle, as a function of time.
The agreement of the equilibrium values with theory is again good although we note that convergence to equilibrium can be slow for large systems and lower temperatures, particularly for the mean energy.  The reason is that when particles are close to each other, their energies are low and MC moves that increase the energy are unlikely to be accepted -- also, the probability of proposing a move into a state where the particles are extremely close is small, so these states are rather hard to access.  This effect is particularly apparent for the mean energy since that quantity is dominated by the smallest gaps in the system, in contrast to the mean gap $\overline{Y}_{\rm g}$, which is dominated by large gaps.  Convergence to equilibrium could presumably be improved by using different MC moves (either with smaller $a_{\rm max}$, or a non-trivial distribution of MC move sizes, or MC moves that move clusters of particles collectively~\cite{whitelam11-molsim}).

\subsection{Thermodynamic limit: large $N,L$ at fixed $\rho,\beta$}
\label{sec:thermo}

So far we considered results for systems with finite numbers of particles.  Here, we briefly discuss the thermodynamic limit in which the temperature $T$ is fixed, and $N,L\to\infty$ with fixed $\rho=N/L$.  It is useful to consider the distribution of a single gap which from (\ref{equ:p1y-finite}) can be written as
\begin{equation}
P_1(y) = \frac{1}{C_0} y^{-\beta} (1-\rho y/N)^{(1-\beta)(N-1)-1},
\end{equation}
where $C_0$ is a normalisation constant.
Using the standard result $\lim_{M\to\infty} (1-x/M)^M = {\rm e}^{-x}$ with $\beta<1$ and $M=N(1-\beta)$ we obtain a Gamma distribution:
\begin{equation}
P_1^\infty(y) = \lim_{N\to\infty} P_1(y) %= \lim_{M\to\infty} \frac{1}{C_L} y^{-\beta} (1-\rho y(1-\beta)/M)^{M} 
= \frac{1}{C_\infty} y^{-\beta} {\rm e}^{-\rho y(1-\beta)} ,
\label{equ:p1inf}
\end{equation}
with $C_\infty=\Gamma(1-\beta)/[\rho(1-\beta)]^{1-\beta}$.
For $\beta=0$ (non-interacting particles), we recover a simple exponential distribution with mean $1/\rho$, as expected.  For $\beta>0$, small gaps are favoured due to the factor $y^{-\beta}$.  At the same time the interaction also enhances the statistical weight of \emph{large} gaps, via the decaying exponential term in (\ref{equ:p1inf}) -- this ensures that the mean gap size remains constant at $1/\rho$, as required.

We also have
\begin{equation}
P_{\rm g}^\infty(y) = \frac{\rho}{C_\infty} y^{1-\beta} {\rm e}^{-\rho y(1-\beta)} ,
\label{equ:pginf}
\end{equation}
from which we see (as expected) that a randomly chosen point in the system is almost surely contained in a gap of size of order $\rho^{-1}$, which remains constant in the thermodynamic limit.  Also the mean energy per site (\ref{equ:enN}) converges to
\begin{equation}
E^\infty = \psi(1-\beta) - \log \rho(1-\beta) ,
\end{equation}
where we used $\psi(x)-\log(x)\to0$ as $x\to\infty$.

\subsection{Behaviour as $\beta\to1^-$ in a finite system}
\label{sec:beta1}

We have explained that $\beta=1$ corresponds to a special temperature for the model, in that the Boltzmann distribution is not normalised at lower temperatures ($\beta>1$).  It is useful to consider briefly the limit $\beta\to 1^-$, in a finite system.  From (\ref{equ:Yg}), one sees that on choosing a random point, the mean size of the gap containing that point is $\overline{Y}_{\rm g}\to L$ as $\beta\to 1^-$.  Since all gaps must be smaller than $L$, this means that as $\beta\to 1^-$ the whole of the system becomes dominated by a single gap, with all the particles located in a single cluster (and separated by much smaller gaps).  More precisely, $\overline{Y}_{\rm g} = L[1-(N-1)(1-\beta)]+O(1-\beta)^2$, from which one sees that the $N-1$ small gaps have an average size of roughly $L(1-\beta)$ as the limit is approached.

This situation, where a single gap occupies almost all of the system, corresponds to a particular kind of condensation phenomenon, as discussed in the next section.  We also note that a similar singularity appears in the trap model of glassy dynamics proposed by Bouchaud~\cite{bouchaud92,monthus96}, for which the partition function is not normalisable for low temperatures: in that case there is an associated stochastic dynamics that is well defined for all $T$ but the system never equilibrates for $T<1$, leading to aging behaviour.  In Section~\ref{sec:multi} below, we investigate the behaviour of our model for $\beta\approx 1$.  However, before embarking on that analysis, we connect the results obtained thus far to previous work on condensation processes.

%To end this section, it is useful to note that for $\beta\to1^-$, the distribution $P_{1}^\infty$ has is dominated by small gaps ($y\to0$) but $P_{\rm g}^\infty$ is dominated by large gaps ($y\to\infty$).  This means that a single cluster 
%
%Taking $\beta\to1^-$ in a finite system, we expect to see a single cluster that contains all the particles in a vanishingly small region of the system, so that the $P_{\rm g}$ is dominated by a single large gap of size $L$.  In the mass transport (chipping) mode, this is a particular type of condensation transition for the mass transport model, as considered in ?? WHAT ??.  \emph{Note: this condensation is driven by the small-$y$ tail of the gap distribution, it's not like the usual Evans situation~\cite{majumdar05} where all is controlled by the large-$y$ tail.}

\subsection{Relation to a chipping process, and to condensation phenomena}
\label{subsec:chip}

As discussed in~\cite{cates98,evans-review}, particle hopping models of the type considered here are related to mass transport models.  To see this, we use the regularisation in which particles cannot approach each other more closely than $\epsilon$, and consider the MC time evolution given above with $a_{\rm max}<\epsilon$.  In this case particles may not overtake each other, so we can order their positions so that $y_i=x_{i+1}-x_i$ (modulo periodic boundaries).  Now define a mass transport model that consists of a periodic lattice of $N$ sites, with mass $y_i$ on each site, so that the total mass is ${\cal M}=\sum_i y_i=L$.  Implementing the particle dynamics for the original model corresponds to a dynamical evolution for the masses $y_i$: if particle $i$ in the original model moves to the right by a distance $a$, this corresponds to a transfer of mass $a$ in the lattice model, from site $i+1$ to site $i$~\cite{cates98}.  The rate for such events depends on the mass transfer $a$ and on the original masses $y_i,y_{i+1}$.  This corresponds to a particular chipping kernel for the mass transport~\cite{majumdar05}.
{This allows the model considered here to be mapped exactly to a mass-transport model whose steady state distribution has the product structure shown in (\ref{equ:y2}).  Such models have been of considerable recent interest -- the masses $y_i$ must be positive but they may be either integer-valued (as in zero-range processes) or real-valued (as in chipping processes or the Brownian energy process)~\cite{cates98,chleboun14,majumdar05,gross11}.}

{For the mass-transport model corresponding to our discussion here, the rates for mass transport in each direction are symmetric: if the probability of moving mass $a$ from $i$ to $i+1$ is $r_{i}(a|y_i,y_{i+1})$ then the probability of transporting the same mass from $i+1$ to $i$ is $\ell_{i+1}(a|y_i,y_{i+1})=r_{i}(a|y_{i+1},y_{i})$.  This ensures that there is no preference in the direction of mass transport.
In other cases~\cite{chleboun14,majumdar05} one instead considers asymmetric models in which mass transport is possible only in one direction, or in which the rates encode a preference for hopping in one particular direction.  Such models can be constructed with steady state probability distributions of product form, as in (\ref{equ:y2}), so clustering and condensation phenomena can be observed in non-equilbrium (asymmetric) systems as well as in equilibrium~\cite{chleboun14}.  Non-equilibrium models with the distribution (\ref{equ:y2}) can be defined and will lead to the same cluster-formation properties discussed here.}

The phenomenon of \emph{condensation} in this mass transport model happens in the thermodynamic limit $N,{\cal M}\to\infty$ at fixed $\phi=N/{\cal M}$: condensation means that a finite fraction of the total mass becomes concentrated on a single site $i$.  {(This may happen either for integer-valued or real-valued masses $y_i$.)}  In the particle model, this corresponds to a situation where one of the gaps between particles takes up a finite fraction of the system.  A large body of previous work~\cite{chleboun14,majumdar05} has considered distributions of the form (\ref{equ:y2}), but with some regularisation at small $y_i$ so that the power law $y^{-\beta}$ might be replaced (for example) by $1/(1+y^\beta)$.  In this case, condensation is generally expected for $\beta>2$~\cite{majumdar05}.  In the thermodynamic limit of the particle model, this means that a single large gap would occupy a finite fraction of the system, with other gaps having typical sizes of order $\rho^{-1}$.

We note that this \emph{condensation} is not generally the same as cluster formation in the model considered here.  Condensation corresponds to a single large gap taking up a finite fraction of the system -- it is a feature associated with a very large gap. Here, cluster formation will be associated with a large number of particles concentrating at a single point -- it is associated with many very small gaps.  A corollary of these small gaps is that some larger gaps must also appear (since the mean gap size is fixed).  In the model considered here, these large gaps take up a finite fraction of the system, as in condensation, but cluster formation need not be linked to this effect.  {(For example, suppose that a single cluster contains half of the particles, with the remainder distributed at random throughout the system.  In that case, there would be no macroscopic gaps but there would be a macrosopic cluster.)}

Moreover, the origin of the condensation behaviour considered here is different from the classical case~\cite{majumdar05}, due to the singular behaviour of (\ref{equ:y2}) for small $y_i$.  As a result, the clustering instability considered here is already present for $\beta>1$ (but only if the regularisation parameter $\epsilon\to0$), while the condensation instability sets in only for $\beta>2$ (and occurs even if $\epsilon>0$).  The instability considered here also has a condensate that contains \emph{all} of the mass in the system, reminiscent of inclusion processes~\cite{gross11}.  {Note that some regularisation of the power laws in (\ref{equ:y2}) is absolutely necessary in models with integer-valued masses since there should be a finite probability $y_i$ is zero in that case.  Hence, since the behavior considered relies on the absence of any regularisation, it must be linked to some extent to the use of continuous masses or, equivalently, the continuous positions $x_i\in[0,L)$ used in the original model definition.}

\section{Limit of multiple clusters}% large $N$, fixed $L=1$ and $\beta\to1$ as $1+b/N$}
\label{sec:multi}

We now to turn to the regime of primary interest for this model.  Inspired by~\cite{max}, we consider a kind of hydrodynamic limit.  To motivate this, fix the density $\rho=N/L$ and increase the system size $L$, but imagine observing the system on a length scale $\ell\sim L$ that is also increasing.  The usual expectation is that as we observe the system on these large scales, a description in terms of individual particles can be replaced by a description in terms of a smooth density profile, as happens (for example) when the motion of a fluid is described by the Navier-Stokes equation.

Mathematically, the limit of large observation scale $\ell$ can be investigated by rescaling particle positions from $[0,L)$ into the unit interval $[0,1)$, defining $\hat{x}_i=x_i/L$, and observing this rescaled system on a length scale $\hat{\ell}=\ell/L$.  Taking this limit at a fixed density, the \emph{mean} spacing between particles in the rescaled system is $1/\hat{\rho}=1/(\rho L)$, which tends to zero~\cite{bertini}.  Assuming that \emph{all} particle spacings tend to zero in this way, the system can be defined in terms of a smooth density profile: in an observation window of size $\ell=\hat{\ell}L$  one expects of order $(\rho \hat{\ell} L)$ particles, which diverges in the hydrodynamic limit.  In this case, one expects a law of large numbers to apply, so that even if the particle positions are random, the fraction of particles in any such region will converge (almost surely) to a deterministic value of order unity. 

To make this argument precise, define the empirical density 
\begin{equation}
\mu(\hat{x}) = (1/N)\sum_i \delta(\hat{x}-\hat{x}_i) ,
\label{equ:mu}
\end{equation}
 where we recall that the $\hat{x}_i$ are the positions of the particles, rescaled into the unit interval.  Since the empirical density is a sum of Dirac delta functions, we clearly cannot expect pointwise convergence to any smooth density profile.  Instead, we consider a weaker form of convergence: the empirical density $\mu$ converges to a smooth density profile $\mu_0$ if for any sufficiently well-behaved test function $f$, one has (almost surely) that $\lim_{N\to\infty} \int_0^1 \mu(x) f(x) \mathrm{d}x =\int_0^1 \mu_0(x) f(x) \mathrm{d}x$.  For example, in the model considered here at equilibrium for $\beta<1$, the statement is true with $\mu_0(\hat{x})=1$, independent of $\hat{x}$.  A more interesting setting for the same question would be: if the system is prepared away from equilibrium with a density profile $\mu_0$ that is not constant, how does this smooth density evolve with time?  For $\beta<1$, we expect some kind of (deterministic) diffusion equation, perhaps with a density-dependent diffusion constant.

However, in this section, we concentrate on a different situation~\cite{max}, in which the empirical measure does not converge to any kind of smooth profile, even as $N\to\infty$.

\subsection{Emergence of clusters}

To achieve this, we modify (increase) the interaction strength as we increase the number of particles, by taking
\begin{equation}
\beta = \frac{N-b}{N-1} 
\label{equ:beta-scaling}
\end{equation}
for some constant $b> 1$, so that $(1-\beta)(N-1)=b-1$.  We have $\beta\to1^-$ as $N\to\infty$, and since $\beta=1$ is the limit of stability of the model, one may expect to see non-trivial behaviour in this limit.  A similar construction was used to define interacting particle systems with multiple condensates~\cite{evans-multi}, and in models with discontinuous condensation~\cite{gross08}.

Now consider an equilibrium configuration of the model, and a random point within the system.  We take the hydrodynamic limit $N\to\infty$ with (\ref{equ:beta-scaling}) and we consider the probability that the random point lies in a gap of size $\hat{y}$, which follows from (\ref{equ:p1y-finite}) and (\ref{equ:pg}), yielding
\begin{equation}
\fl
\hat{P}_{\rm g}(\hat{y}) = \lim_{N\to\infty} LP_{\rm g}(\hat{y}L) = \lim_{N\to\infty} \left[  \frac{\rho L^{b-1}(\hat{y}L)^{{\frac{b-1}{N-1}}}}{C_1} (1-\hat{y})^{b-2} \right] = (b-1) (1-\hat{y})^{b-2} ,
\label{equ:Phat-y}
\end{equation}
where we used $B(u,v)\approx (1/u)$ for small $u$ in order to obtain the limiting behaviour of $C_1$.
The key point is that this limiting probability density exists for all $\hat{y}>0$, is independent of $L$, and is normalised to unity.  Recall that $\hat{y}$ is the size of a gap between two particles, scaled by the system size.  Hence (\ref{equ:Phat-y}) means that a randomly chosen point in an equilibrium configuration lies (almost surely) in a gap whose size is comparable with the system size.  This is not at all the case in the conventional thermodynamic limit, in which almost all gaps are comparable with the inverse density $(1/\rho)=L/N$. In that case all of the scaled gaps $\hat{y}_i=y_i/L$ tend almost surely to zero as we take $N\to\infty$, so the limiting probability density $\hat{P}_{\rm g}(\hat{y})$ would be concentrated entirely at $\hat{y}=0$.

Finally we note that the energy per gap  in (\ref{equ:enN}) diverges in this hydrodynamic limit as
\begin{equation}
\frac{1}{NJ}\langle E\rangle \simeq \frac{-N}{b-1} + \log N + c ,
\end{equation}
with $c$ of order unity as $N\to\infty$.  (We used $\psi(x)\simeq -1/x$ as $x\to0$.)  Since this quantity is equal to $\langle 
\log y_i\rangle$, we conclude that the average must be dominated by exponentially small gaps, with $y_i \lesssim {\rm e}^{-N/(b-1)}$, consistent with the idea that the clusters of particles concentrate on single points, in the limit.

\begin{figure}
\hspace{3cm}
\includegraphics[width=11cm]{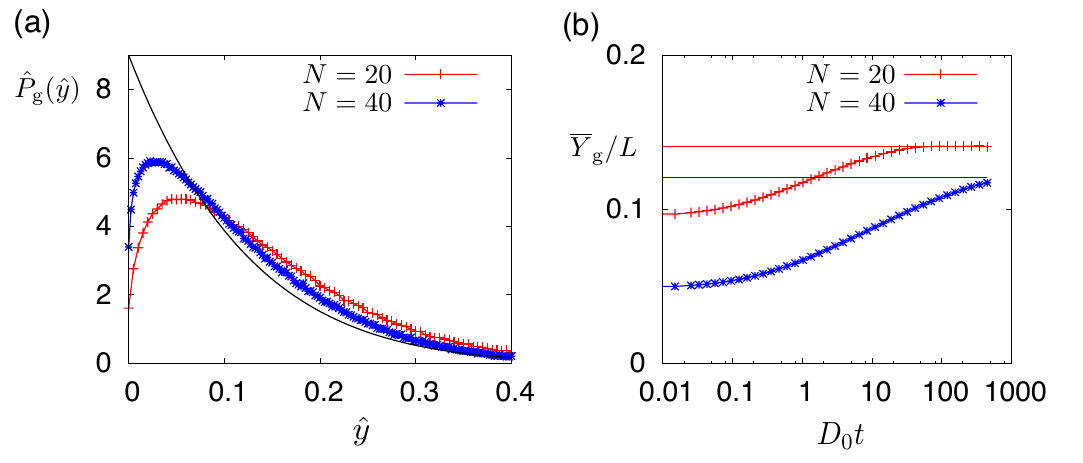}
\caption{Numerical results illustrating the limit where multiple clusters appear.  (a)~Distribution of the scaled gap size $\hat{P}_{\rm g}(\hat{y})$, for two different system sizes, as one takes the hydrodynamic limit according to (\ref{equ:beta-scaling}) with $b=10$. The limiting distribution (\ref{equ:Phat-y}) is shown as a solid line. (b) Convergence to equilibrium of the mean (rescaled) gap size $\overline{Y}_{\rm g}/L$; the ($L$-dependent) equilibrium values for this quantity are indicated.  In the hydrodynamic limit, this average value approaches $1/b=0.1$.  The time taken to converge to equilbrium increases rapidly as $N$ increases, primarily because the interparticle attractions are becoming stronger, in accordance with (\ref{equ:beta-scaling}).}
\label{fig:clust}
\end{figure}

In Fig.~\ref{fig:clust} we compare our numerical results to the theoretical predictions of this section: we illustrate the convergence of $\hat{P}_{\rm g}(\hat{y})$ to its limiting form  as $N\to\infty$, and the convergence with time of the (rescaled) mean gap size $\overline{Y}_{\rm g}/L=\int \hat{y} \hat{P}_{\rm g}(\hat{y}) \mathrm{d}\hat{y}$.  As in Fig.~\ref{fig:finiteN}, the convergence with respect to time $t$ is rather slow when interactions are strong and systems are large, but these results are sufficient to illustrate our main conclusions.  (Note also, the presence of exponentially small gaps means that numerical precision will limit our ability to resolve the fine detail in this problem when $N$ is large and {attractions are strong}.)

The physical interpretation of this result is that the strong attractive interactions between particles lead to the formation of clusters (recall Sec.~\ref{sec:beta1}).  Within a cluster there are many small gaps between particles, but these gaps are so small that a point picked at random has probability zero of being in such a gap.  Between the clusters, there are large gaps, whose sizes are comparable with the system.  These are the gaps that contribute to (\ref{equ:Phat-y}).

If we consider the empirical density $\mu(\hat{x})$ defined in (\ref{equ:mu}), the fact that the hydrodynamic limit consists of clusters separated by large gaps means that $\mu$ does not converge to any smooth profile $\mu_0$.  Rather, assuming that a hydrodynamic description exists, we should think that $\mu$, which is a sum of $N$ Dirac delta functions, should converge (as $N\to\infty$) to some $\mu_0(\hat{x})=\frac{1}{n} \sum_j M_j \delta(\hat{x}-\hat{X}_j)$ where $M_j$ is the mass of the $j$th cluster and $\hat{X}_j$ is its position.  Clearly the number of clusters $n\ll N$.  Moreoever, as the particle model evolves in time, one cannot describe the time evolution of the corresponding $\mu_0$ by any kind of deterministic diffusion equation.  Instead it should solve some kind of stochastic partial differential equation that can describe the random motion of the clusters in the system.

\begin{figure}
\includegraphics[width=15cm]{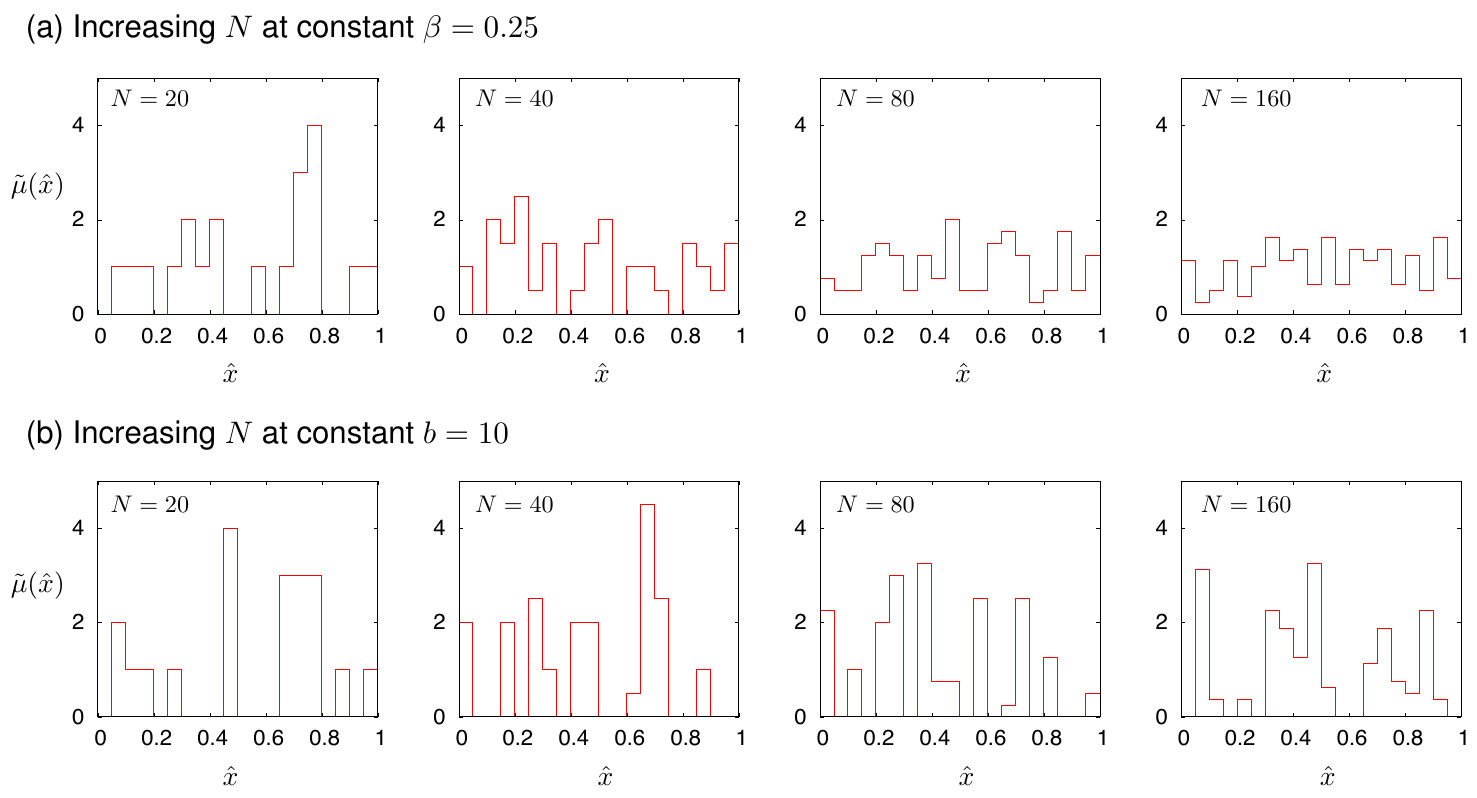}
\caption{Illustration of the hydrodynamic limit. We show histograms of the density based on configurations of the system for various $N$.  [To define $\tilde\mu(\hat{x})$, we divide the unit interval into bins (subintervals) $B_n=[(n-1)\delta \hat{x},n\delta \hat{x})$, for integer $n$. We take $\delta x = 0.05$ so $1\leq n \leq 20$.  Then $\tilde\mu(\hat{x})$ is the density of particles in the bin that contains the point $\hat{x}$.] (a) Increasing $N$ at constant $\beta=0.25$: as the number of particles increases then the local density is self-averaging and $\tilde{\mu}$ converges to a flat profile, as expected for a diffusive system at equilibrium.  (That is, the density satisfies a law of large numbers within each bin.) (b) Increasing $N$ and varying $\beta$ according to (\ref{equ:beta-scaling}) with $b=10$ does not lead to a smooth density profile: multiple clusters of particles persist even as $N\to\infty$ because the particle correlations are so strong there is no law of large numbers within each bin.  (For $N=160$ and $b=10$, the limitations of our numerical method mean that the system may not be completely converged to equilibrium, but the data are sufficient to illustrate the qualitative behaviour.)}
\label{fig:zoom}
\end{figure}

The resulting situation is illustrated numerically in Fig.~\ref{fig:zoom}, where we represent the density $\mu$ by histograms, for various different configurations.  Taking $N\to\infty$ at fixed $\beta=0.25$, the results are consistent with convergence to a smooth profile: there are sufficiently many particles in each region of space that a law of large  numbers applies so the local (smoothed) density $\tilde\mu\to1$ as $N\to\infty$.
However, if we fix $b$ in (\ref{equ:beta-scaling}) and take the limit $N\to\infty$, the results indicate that there is no convergence to a smooth density profile: {the variation in the density between the bins is of the same order as the density itself.}

The emergence of several (or many) clusters in this model raises several interesting questions.  In the remainder of this work, we investigate how many of these clusters there are and how they are distributed in space.  Other possible questions, such as the dynamics of these clusters~\cite{godreche05}, are beyond the scope of this work, but we discuss them briefly in Sec.~\ref{sec:conc}.

\subsection{Statistics of clusters}

From (\ref{equ:Phat-y}), one sees that on choosing a random point in the system, the average size of the gap containing this point is $\overline{Y}_{\rm g}/L=(1/b)$.  From this result, one might suppose that there are typically $b$ clusters within the system, separated by gaps of this typical size.  In fact the situation is rather more complicated.  

To see this, suppose that we choose two random points in the system.  For a finite system with $N$ particles and interaction parameter $\beta$, we have a joint probability density for the two gap sizes
\begin{equation}
\fl
P_2({y},{y}') = P_{\rm g}({y}|N,L) \cdot (y/L) \delta({y}-{y}')  + {P_{\rm g}({y}|N,L) \cdot (1-y/L) P_{\rm g}({y}'|N-1,L-{y})} ,
\label{equ:p2yyp}
\end{equation}
where $P_{\rm g}(y|N,L)$ is the distribution (\ref{equ:pg}) for a system of size $L$ containing $N$ particles.  The first term in (\ref{equ:p2yyp}) accounts for the case where both points are in the same gap, while the second is the case where they are in different gaps.  (If the first point to be chosen is in a gap of size ${y}$, the probabilities of these two outcomes are ${y}/L$ and $1-{y}/L$ respectively.)  In the case where the two points are in different gaps, we have used the fact that gaps are independent, so once the first gap is fixed, the distribution of the second gap is obtained by considering an equivalent system with size $L-y$, and with one fewer particle.
If we now take the hydrodynamic limit according to (\ref{equ:beta-scaling}), we define $\hat{P}_2(\hat{y},\hat{y}')= \lim_{N\to\infty} L^2 P_2(\hat{y}L,\hat{y}'L)$ and obtain
\begin{equation}
\fl
\hat{P}_2(\hat{y},\hat{y}') = (b-1) \hat{y} (1-\hat{y})^{b-2} \delta(\hat{y}-\hat{y}') +  (b-1)^2 (1-\hat{y}-\hat{y}')^{b-2} \Theta(1-\hat{y}-\hat{y}') ,
\label{equ:p2yyp-inf}
\end{equation}
where the step function $\Theta$ in the second term enforces that the sum of the two gaps must be less than the system size.  Note that this distribution is symmetric in $\hat{y},\hat{y}'$ (as it should be).

From the second term in (\ref{equ:p2yyp-inf}) we see when the two points are located in different gaps, then \emph{both} of these gaps almost surely have sizes comparable with the whole system.  If we condition on this case, we can consider the distribution of the second gap $\hat{y}'$, given a particular value of the first gap $\hat{y}$.   Rescaling the size of the second gap as $\hat{y}''=\hat{y}'/(1-\hat{y})$, the distribution of $\hat{y}''$ is exactly the original $\hat{P}_{\rm g}$.  (This fact is true only in the hydrodynamic limit $N\to\infty$ because it relies on the fact that replacing $N-1$ by $N$ in the last term of (\ref{equ:p2yyp}) has no effect on the limiting behaviour.)  Analysing the joint distribution of three or more gaps is a straightforward extension of the same procedure.  The conclusion is that if we remove a single large gap from the system, the distribution of the remaining large gaps in the remainder of the system is the same (up to rescaling) as the distribution of all the gaps within the whole system.  Hence we conclude that there are in fact infinitely many large gaps, and hence \emph{infinitely many} clusters in the system, arranged in a hierarchical structure.
%Indeed, if we condition on the two ponts being in different gaps, the distribution of the second gap $\hat{y}'$ is the same as $\hat{P}_{\rm g}(\hat{y})$, except 
%Moreover, the extension of this result to three or more points is straightforward.  For any number of points, there is a non-zero probability that \emph{all} points lie in different gaps, and if we condition on this case, \emph{all} these gaps almost surely have sizes of the order of the system size.  From this we conclude that there are in fact infinitely many large gaps, and hence \emph{infinitely many} clusters in the system, arranged in a hierarchical structure.  Physically, if we remove any one gap and consider the remaining particles distributed through the remainder of the system, the resulting system is statistically identical to a rescaled version of the original system.  

\begin{figure}
\hspace{2.5cm}\includegraphics[width=10cm]{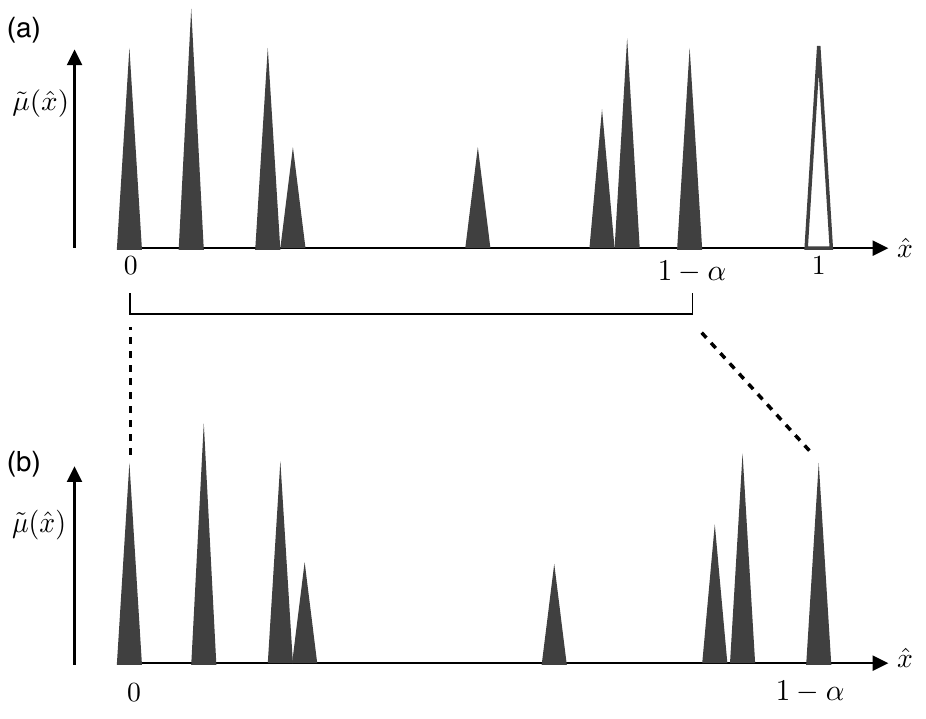}
\caption{Illustration of the statistics of the clusters within the system.  (a) Sketch of the (smoothed) empirical density for a configuration containing several clusters, within the unit interval.  Each cluster is depicted as a triangle with finite width: there are many particles within each cluster but as $N\to\infty$ each cluster should concentrate on a single point.  The origin $\hat{x}=0$ has been placed at the centre of a cluster and the periodic image of this cluster is shown at $\hat{x}=1$ by an unfilled triangle.  (b) If we ignore the last gap (which has size $\alpha$) and zoom in on the remainder of the system, the distribution of clusters within this subsystem is statistically identical (as $N\to\infty$) to their distribution in the full system.}
\label{fig:clust-sketch}
\end{figure}

A sketch of this situation is shown in Fig.~\ref{fig:clust-sketch}, illustrating how the system includes gaps of all sizes, arranged in a fractal (self-similar) structure.  Nevertheless, we emphasise that since the number of particles in the system has already been taken to infinity, typical gaps between particles are vanishingly small on the scale shown here.  So while the number of the clusters in the system is infinite, each cluster contains a very large (presumably infinite) number of particles.

\subsection{The limit $b\to1^+$}

{Note that we must have $b>1$ in (\ref{equ:beta-scaling}) since otherwise $\beta>1$ and the whole analysis breaks down (all our arguments start from (\ref{equ:y2}) which requires $\beta<1$).  However, the limit $b\to 1$ is of interest.  In this case, the hierarchical structure discussed in this section simplifies: the physical picture is that on choosing a random point, the gap containing this point covers almost all of the system, up to corrections that vanish as $N\to\infty$.
Mathematically, it is easy to verify that 
\begin{equation}
\lim_{\Delta\searrow 0}\; \lim_{b\searrow 1}\; \int_{1-\Delta}^1 \hat{P}_{\rm g}(\hat{y}) \mathrm{d}\hat{y} = 1,
\label{equ:y1as}
\end{equation}
so that $\hat{y}\to1$, almost surely.  Hence one may write for $b\to1$ that $\hat{P}_{\rm g} \to \delta(1-\hat{y})$ and (\ref{equ:p2yyp-inf}) simplifies to $\hat{P}_2(\hat{y},\hat{y}') = \delta(\hat{y}-\hat{y}') \delta(1-\hat{y})$.  That is, on choosing two points in the system, they almost surely lie in the same large gap, which covers (almost) the whole system.  In this case, all of the particles are concentrated in a single cluster.
This is the same situation as was discussed informally in Sec.~\ref{sec:beta1} where $\beta\to1$ at fixed $N$. Writing $\beta=1-\frac{b-1}{N-1}$, one sees that taking $b\to1^+$ at fixed $N$ (as in Sec.~\ref{sec:beta1}) has the same result as taking $N\to\infty$ and then $b\to 1^+$.  

More generally, one may take the joint limit $\beta\to1, N\to\infty$ with $1-\beta=cN^{-\alpha}$ for any $\alpha\geq 0$ and $c>0$.  We expect all particles in a single cluster for $\alpha>1$ (which includes the case discussed in Sec.~\ref{sec:beta1}); for $\alpha<1$ we expect $\hat{y}\to0$ almost surely so there are no macroscopic clusters (this case includes the thermodynamic limit discussed in Sec.~\ref{sec:thermo}, which corresponds to $\alpha=0$).  For $\alpha=1$ one has a hierarchy of clusters as discussed in this section, but one recovers the single cluster on taking $c\to0$.}

\section{Conclusion and outlook}
\label{sec:conc}

We have defined a model of interacting particles on the real line, which has an instability at temperature $T^*=J$.  Below this temperature, particles attract each other so strongly that the gaps between adjacent particles tend to zero, and the system is unstable to collapse at a single point.  However, the system is well-behaved for $T>T^*$: particles attract each other and assemble into clusters. All clusters are finite in the thermodynamic limit, and the system has a hydrodynamic limit in which the macroscopic density is smooth.  We have shown that if we take a hydrodynamic limit in which $T\to T^*$ from above as the number of particles tends to infinity, this system has a well-defined equilibrium state in which density profiles are not at all smooth: instead particles self-organise into clusters that are arranged in a self-similar hierarchical structure.  

The limiting process that we took in order to arrive at this situation was somewhat unusual, but similar methods have been used in zero-range processes~\cite{evans-multi,gross08}.  Our analysis of this model further accentuates the rich phenomenology that is accessible even in deceptively simple interacting particle systems.  It also raises several interesting new questions.

Our model was inspired by~\cite{max}, in which a similar model was proposed, with the same invariant measure (compare the first equation in section 2 of that paper with our Eq.~(\ref{equ:y2}), and note that $\beta$ in that work corresponds to our $b-1$, up to corrections of order $1/N$).  The results of that work indicate that this model has an underlying (abstract) geometrical structure related to the Wasserstein metric -- this is a metric in the space of density profiles $\mu$, with connections to diffusive processes: see~\cite{jz} for a physical discussion and~\cite{adams} for a more mathematical presentation.  From the results presented here, the precise connection between this geometrical structure and our model is not clear to us: we expect that the motion of the clusters that form in the system should be described by a stochastic process in this abstract space, but this requires further investigation.  
{A more rigorous analysis of the steady state of the model would also be useful, particularly regarding the limiting behaviour of the empirical measure $\mu(x)=N^{-1}\sum_i \delta(x-x_i)$ as $N\to\infty$ at fixed $b>1$.  We have argued that this measure consists of clusters separated by large gaps, and we have argued that the gaps are independently distributed.  However, the distribution of the number of particles in each cluster is not available from the analysis performed here, so a full characterisation of the limiting measure requires more detailed investigation.}

Independent of those questions, it would also be very interesting to characterise the motion of the clusters of particles that form in this system, when the hydrodynamic limit is taken.  In particular, we expect the clusters to move diffusively~\cite{godreche05}, and they should presumably undergo fusion and fission processes when they encounter one another.  Certainly, the Langevin dynamics (\ref{equ:lang}) imply that the centre of mass of a cluster of mass $N$ moves with a diffusion constant of order $1/N$, but other processes in the system may also be relevant (for example exchange of particles between clusters, which can even lead to cluster evaporation).  Also, the MC dynamics defined here are different in general from (\ref{equ:lang}), given that we take $N\to\infty$ and $\beta\to1$ at fixed $a_{\rm max}$.  Further numerical or analytical results for these processes would be valuable, either for this system or for other systems where multiple clusters (or condensates) appear in large systems~\cite{evans-multi}.  We hope to revisit these questions in a later work.

\ack RLJ thanks Max von Renesse and Johannes Zimmer for helpful discussions, which motivated this work.

\end{document}